\documentclass[fleqn,usenatbib]{mnras} 

\usepackage{newtxtext,newtxmath}

\usepackage[T1]{fontenc}
\usepackage{ae,aecompl}

\usepackage{graphicx}	
\usepackage{amsmath}	
\usepackage{amssymb}	

\title[Periodicities in ULXs, HMXBs and DPVs]{Orbital and super-orbital periods in ULX pulsars, disc-fed HMXBs, Be/X-ray binaries and double-periodic variables}

\author[Townsend \& Charles]{L. J. Townsend$^{1,2,3}$\thanks{E-mail: lee@saao.ac.za (LJT)}
\& P. A. Charles$^{4}$
\\
$^{1}$South African Astronomical Observatory, P.O. Box 9, Observatory, 7935, Cape Town, South Africa\\
$^{2}$Southern African Large Telescope, P.O. Box 9, Observatory, 7935, Cape Town, South Africa\\
$^{3}$Department of Astronomy, University of Cape Town, Private Bag X3, Rondebosch, 7701, South Africa\\
$^{4}$Department of Physics \& Astronomy, University of Southampton, Highfield, Southampton SO17 1BJ, UK\\
}
\date{Accepted 2020 April 26. Received 2020 April 08; in original form 2020 February 20}

\pubyear{2020}

\begin{document}
\label{firstpage}
\pagerange{\pageref{firstpage}--\pageref{lastpage}}
\maketitle

\begin{abstract}
We present evidence for a simple linear relationship between the orbital period and super-orbital period in ultra-luminous X-ray (ULX) pulsars, akin to what is seen in the population of disc-fed neutron star super-giant X-ray binary and Be/X-ray binary systems. We argue that the most likely cause of this relationship is the modulation of precessing hot spots or density waves in an accretion or circumstellar disc by the binary motion of the system, implying a physical link between ULX pulsars and high-mass X-ray binary (HMXB) pulsars. This hypothesis is supported by recent studies of Galactic and Magellanic Cloud HMXBs accreting at super-Eddington rates, and the position of ULX pulsars on the spin period--orbital period diagram of HMXBs. An interesting secondary relationship discovered in this work is the apparent connection between disc-fed HMXBs, ULXs and a seemingly unrelated group of early-type binaries showing so-called ``double-periodic'' variability. We suggest that these systems are good candidates to be the direct progenitors of Be/X-ray binaries.
\end{abstract}




\section{Introduction}

X-ray pulsars have been known to exist in high-mass X-ray binaries (HMXB) for $\sim$50 years. The neutron star producing these X-rays accretes either from a circumstellar disk of material around a late O-type or early B-type star, or via Roche-Lobe overflow or stellar wind of a super-giant companion. The latter of these systems, known as super-giant X-ray binaries (SGXB), account for a small fraction of the known HMXB population and consist of both neutron stars and black holes. The former, known as Be/X-ray binaries (BeXRB), are much more numerous. With well in excess of 250 confirmed systems in the Milky Way and Magellanic Clouds (\citealt{2006A&A...455.1165L}; \citealt{2016A&A...586A..81H}), at least half of which contain a neutron star, BeXRBs make up one of the largest known classes of accreting pulsars. Their observed X-ray luminosity can range between L$_{\mathrm{X}}\sim$ 10$^{32}$ -- 10$^{39}$ erg s$^{-1}$, which can be modulated on the binary period or by the intrinsic variability timescale of the mass-donating Be star.

\begin{table*}
	\centering
	\caption{Spin, orbital and super-orbital periods, donor spectral type and peak X-ray luminosity of the confirmed ULX pulsars and super-Eddington BeXRBs.}
	\label{tab:example_table}
\begin{tabular}{l c c c c c}
\hline
\textbf{System name} & \textbf{P$_{\mathrm{spin}}$} & \textbf{P$_{\mathrm{orb}}$} & \textbf{P$_{\mathrm{sup}}$} & \textbf{Sp. Type} & \textbf{L$_{\mathrm{X, max}}$}\\
 & s & d & d & & 10$^{39}$ erg $^{-1}$\\
\hline
A0538-66$^{1,2,3,4}$ & 0.069 & 16.64 & 421 & B1 IIIe & 0.8\\
NGC7793 P13$^{5,6,7,8}$ & 0.42 & 63.9 & $\sim$1500$^{*}$ & B9Ia & 6\\
NGC5907 ULX-1$^{9,10,11}$ & 1.14 & 5.3 & 78.1 & red giant? & 220\\
M82 X-2$^{12,13,11}$ & 1.37 & 2.5 & $\sim$60 & early type? & 18\\
NGC1313 X-2$^{14}$ & 1.46 & $<$6 & - & early type? & 5\\
M51 ULX-7$^{15,16,17}$ & 2.8 & 2.0 & 38.5 & OB? & 7.1\\
SMC X-3$^{18,19,20}$ & 7.77 & 44.92 & 1116 & B1-1.5 IV-Ve & 1.2\\
Swift J0243.6+6124$^{21,22,23}$ & 9.86 & 27.59 & $\sim$1000$^{**}$ & Be & 1.8\\
NGC300 ULX-1$^{24,25,26}$ & 16.58 & $>$300? & - & RSG & 4.7\\

\hline
\end{tabular}
\begin{flushleft}
\textbf{Note.} M51 ULX-8 also harbours a neutron star (\citealt{2018NatAs...2..312B}), but no spin, orbital or super-orbital periods are known.\\
$^{*}$Beat period suggested by \cite{2018A&A...616A.186F} between the orbital period and 66.9\,d X-ray period.\\
$^{**}$Approximate length of the long-term variation noted by \cite{2017ATel10811....1S} and \cite{2017ATel10989....1N}. See section 4 for details.
\end{flushleft}
\begin{flushleft}
$^{1}$\cite{1982Natur.297..568S}; $^{2}$\cite{1978MNRAS.183P..11W};
$^{3}$\cite{2003MNRAS.339..748M};
$^{4}$\cite{2017MNRAS.464.4133R}; 
$^{5}$\cite{2016ApJ...831L..14F}; $^{6}$\cite{2017MNRAS.466L..48I}; $^{7}$\cite{2018A&A...616A.186F}; $^{8}$\cite{2011AN....332..367M}; 
$^{9}$\cite{2017Sci...355..817I}; $^{10}$\cite{2016ApJ...827L..13W}; 
$^{11}$\cite{2019ApJ...871..231H}; 
$^{12}$\cite{2014Natur.514..202B}; $^{13}$\cite{2019ApJ...873..115B}; 
$^{14}$\cite{2019MNRAS.488L..35S};
$^{15}$\cite{2019arXiv190604791R}; $^{16}$\cite{2019arXiv191204431B}; $^{17}$\cite{2020MNRAS.491.4949V}; 
$^{18}$\cite{2017MNRAS.471.3878T}; $^{19}$\cite{2011MNRAS.413.1600R}; $^{20}$\cite{2008MNRAS.388.1198M}; $^{21}$\cite{2017ATel10809....1K}; $^{22}$\cite{2018ApJ...863....9W}; $^{23}$\cite{2017ATel10822....1K}; $^{24}$\cite{2019MNRAS.488.5225V}; $^{25}$\cite{2018MNRAS.476L..45C}; $^{26}$\cite{2019ApJ...883L..34H};
\end{flushleft}
\end{table*}

Many periods exist in the light curves of BeXRBs. In X-rays, one can often determine the spin period of the neutron star, the orbital period and the super-orbital period(s) of the binary. {\footnote{HMXB super-orbital periods are distinct from those seen in low-mass X-ray binaries and cataclysmic variables where the accreting compact object exceeds the donor mass by at least a factor 3.  This causes the disc to become eccentric and precess, leading to the presence of ``superhumps'' with periods just a few percent greater than P$_{\mathrm{orb}}$ (e.g. \citealt{1991MNRAS.249...25W}).}} Typically observed values for these parameters are: P$_{\mathrm{spin}}\sim$ 1 -- 5000\,s, P$_{\mathrm{orb}}\sim$ 10 -- 1000\,d, P$_{\mathrm{super}}\sim$ 500 -- 10,000\,d. In many cases, one is also able to independently determine the orbital and super-orbital periods from optical light curves. This is because the orbital motion of the neutron star often causes a small optical flare as it passes through the equatorial disc around the rapidly rotating Be star, and because the super-orbital period is caused by the formation and depletion of this circumstellar disc, or rotating density waves therein. \cite{2011MNRAS.413.1600R} studied the OGLE long-term optical light curves of the BeXRB population in the Small Magellanic Cloud (SMC) and found a strong correlation between the orbital and super-orbital periods. The most likely explanation for this is that a precessing density wave or warping of the circumstellar disc around the Be star is being moderated by the extent of the neutron star's orbit.

Such super-orbital periods are also seen in disc accreting SGXBs on slightly shorter timescales than in BeXRBs. These periods are produced in a similar manner, namely through the precession of an asymmetric disc, except this time it is an accretion disc around the neutron star that is being precessed (e.g. Her X-1 and SMC X-1; \citealt{1980ApJ...237..169B}; \citealt{2005ApJ...633.1064H}). In wind-accreting sources (i.e. standard SG systems), on the other hand, the existence of a large accretion disc is unlikely, given the low angular momentum of the accreted matter and the strong magnetic field of the neutron star. However, \cite{2013ApJ...778...45C} discovered super-orbital periods in a number of SG systems, challenging this assumption. So far, no clear explanation for these super-orbital periods has emerged. The formation of a temporary accretion disc is unlikely, due to the clear presence of these features in the long-time monitoring light-curves of the \textit{RXTE} and \textit{Swift} all sky monitors.

Ultra-luminous X-ray sources (ULX) are classified as non-nuclear X-ray sources with an observed X-ray luminosity above $10^{39}$ erg s$^{-1}$. Several hundred of these systems have now been discovered, most of which are confirmed or are thought to host stellar-mass compact objects (see \citealt{2017ARA&A..55..303K} for a recent review). So far, only 7 ULXs have been confirmed to harbour neutron star primaries, either through the detection of pulsations or cyclotron features (see Table \ref{tab:example_table} for details). Super-orbital periods are also observed in ULXs (Table \ref{tab:example_table} and references therein) suggesting that the physical processes occurring in ULXs might be somehow connected to those in standard HMXBs. Recent studies of known HMXBs undergoing super-Eddington outbursts (e.g. SMC X-3, \citealt{2017MNRAS.471.3878T}; Swift J0243.6+6124, \citealt{2018ApJ...863....9W}) lend further support to there being a fundamental link between HMXBs and ULXs.

Since beginning this work, we have become aware of an apparently unrelated group of early-type binaries that display so-called ``double-periodic'' variability (DPV) in their light-curves (\citealt{2003A&A...399L..47M}), on timescales somewhat shorter than the BeXRBs systems. These DPV stars have also been observed as spectroscopic binaries (e.g. \citealt{2005MNRAS.357.1219M}; \citealt{2012MNRAS.427..607M}; \citealt{2015MNRAS.448.1137M}; \citealt{2014A&A...567A.140B}), showing that they consist of an early B-type main sequence star and a hot, lower-mass, but evolved companion (i.e. an Algol-like system). \cite{2003A&A...399L..47M} discuss the population of DPV stars in the Magellanic Clouds and show a highly significant correlation between the two periods. They interpret the shorter period (which covers the range $\sim$2--16d) as the orbital period, and the longer one as being associated with the disc around the accreting B star.  The DPV periods are related by the expression P$_{2}$ = 35.2 $\pm$ 0.8 P$_{1}$. 

In this letter, we examine the correlation between the periodicities in HMXBs, ULXs and DPVs to better understand the connection between these systems. We present a relationship between the orbital and super-orbital periods seen in ULX pulsars that suggests a majority of NS-ULX systems could harbour a massive, early-type companion. Finally, we note that the period correlation seen in DPV systems with a highly evolved primary make them excellent candidates to be the direct progenitors of BeXRB systems. 

\begin{figure*}
	\includegraphics[width=2\columnwidth]{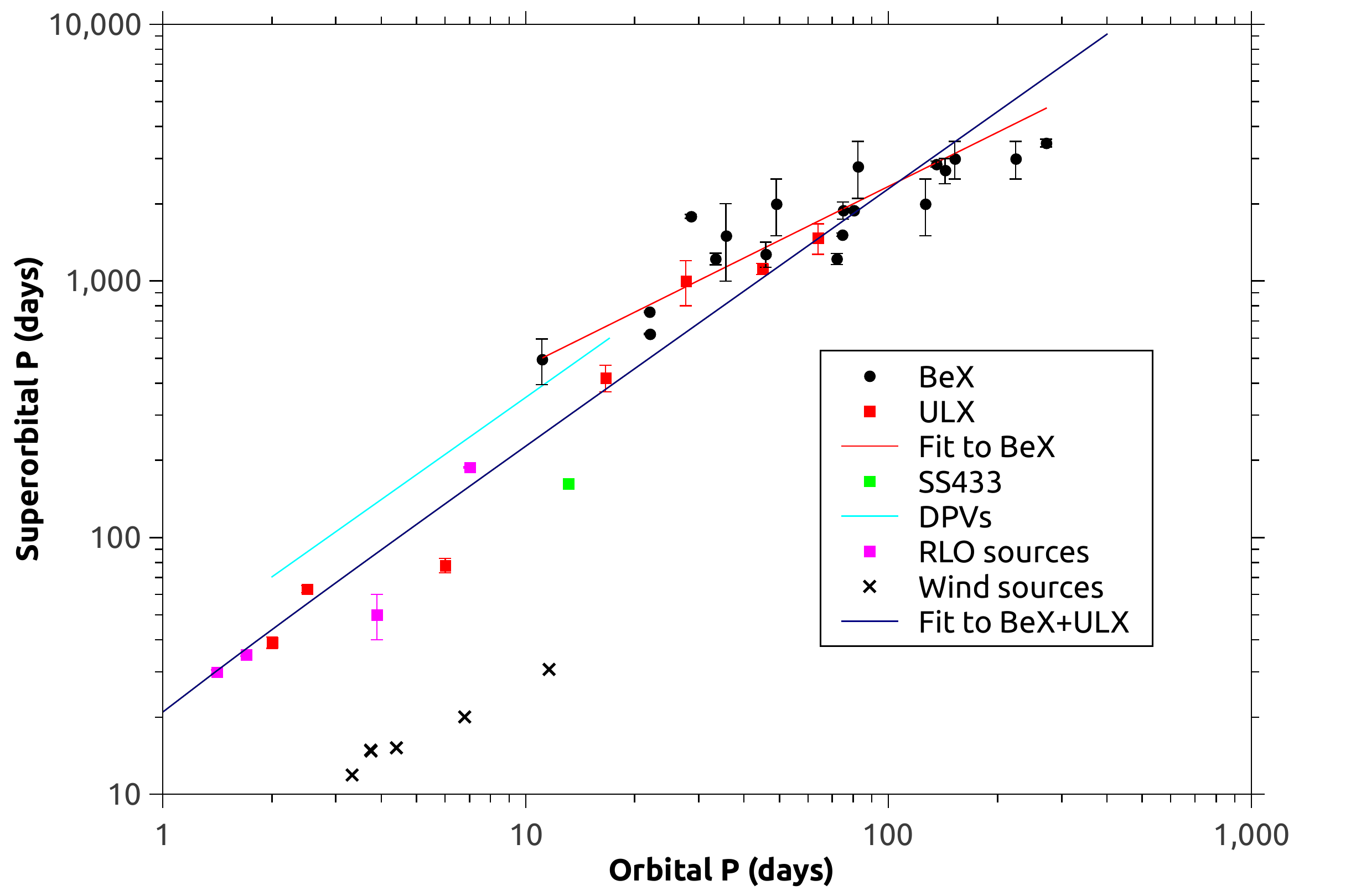}
    \caption{Orbital period vs. super-orbital period for all ULX pulsars (red squares), BeXRBs (black circles), RL-filling SGXBs (pink squares) and wind-fed SGXBs (black crosses) where this information is known. The green square is SS433 for comparison. The red line is the fit to the BeXRBs as shown in Rajoelimanana et al. The pale blue line represents the relationship noted by Mennickent et al. in DPVs. The black line is the combined fit to all ULXs, RL-filling SGXBs and BeXRBs.}
    \label{fig:example_figure}
\end{figure*}

\section{The population of ULX pulsars and super-Eddington HMXBs}

In Table \ref{tab:example_table} we collate details of all periodicities, donor spectral types and peak X-ray luminosities for all ULX pulsars and super-Eddington XRB pulsars currently known. The table includes three BeXRBs that have been observed to accrete well above the Eddington limit for a neutron star, and six ULXs associated with a pulsating X-ray source. One further ULX (M51 ULX-8; \citealt{2018NatAs...2..312B}) has been confirmed to host a neutron star from the detection of cyclotron lines in its X-ray spectrum, bringing the total number of NS-ULXs to seven.

\section{Results}

Figure \ref{fig:example_figure} shows super-orbital against orbital period for all the binary systems discussed so far in this work. Immediately obvious is the correlation between these two periodicities for all systems. We show the fit to the BeXRBs published by \cite{2011MNRAS.413.1600R} as a red line, and the relationship noted by \cite{2003A&A...399L..47M} in DPVs as a pale blue line, both plotted over their original range of applicable periods. A fit through all the ULX and HMXB data (excluding the wind fed systems) is plotted as a black line. The wind fed systems are excluded from this fit due to the uncertainty around the origin of the super-orbital period in these systems. Their placement in Figure \ref{fig:example_figure}, far removed from the trend seen in ULXs, RL overflow SGXBs and BeXRBs, suggests the origin of the periods in these systems is different to the rest of the HMXB and ULX populations, in spite of the slope being similar.

The strong correlation between periods seen in ULXs, RL overflow SGXBs and BeXRBs is shown both in the combined fit to the data and in a simple rank correlation coefficient, with an R$^2$ value of 0.999.  However, we note that there is obviously significant scatter about this line compared to the nominal errors associated with determining the P$_{\mathrm{sup}}$ values, and this could be reduced by fitting a higher order polynomial, but we do not feel this to be justified with the data currently available. A linear fit between the two periods gives P$_{\mathrm{sup}}$ = 22.9 $\pm$ 0.1 P$_{\mathrm{orb}}$, which is slightly lower, but comparable to, the value determined for DPV systems discussed in section 1. 

In the next section, we argue that the correlation in the periods in Figure \ref{fig:example_figure} is strong evidence in support of the idea that many ULX pulsars may be SGXBs or BeXRBs accreting at super-Eddington rates. We also suggest that the similarity between the DPV and HMXB period correlations points toward these systems being progenitors of BeXRBs.

\section{Discussion}

\subsection{Super-Eddington HMXBs}

The X-ray outbursts of A0538-66, SMC X-3 and Swift~J0243.6+6124 are typical of many large outbursts seen in other BeXRB systems, except at a much higher luminosity. There is also no difference in the mechanism by which super-orbital periods are produced in BeXRBs in the sub-Eddington and super-Eddington regimes. These periods are considered to be a result of the precession of a hot spot or density wave or other modulation of the circumstellar disc of the Be star. The precession period is modulated by the orbit of the neutron star, causing the relationship seen in Figure \ref{fig:example_figure}. The same can be said for the RL-filling SGXBs, except in this case it is the precession of the accretion disc around the neutron star that causes the super-orbital periodicity. It is worth highlighting that super-Eddington HMXBs are indistinguishable from ULXs in Figure \ref{fig:example_figure}. \cite{2013ApJ...778...45C} show that there is also a linear correlation between the periods seen in wind-fed SGXBs, though these are very different in behaviour to the RL-filling periods and the cause has not been well established due to the lower probability of an accretion disc forming.

When compiling the data listed in Table \ref{tab:example_table}, we found that no confirmed super-orbital period existed for Swift~J0243.6+6124. However, both \cite{2017ATel10811....1S} and \cite{2017ATel10989....1N} show optical light curves of this source displaying long period variability that suggests a possible P$_{\mathrm{sup}}$  around 1000\,d. Using this value, we plot Swift~J0243.6+6124 in Figure \ref{fig:example_figure} and show that it sits precisely amongst the other BeXRBs, thus confirming the $\sim$1000\,d photometric variation as the true super-orbital period in this system.  Therefore, in spite of the scatter, the relationship in Figure \ref{fig:example_figure} has considerable predictive potential for long-term variability studies.

\subsection{ULX pulsars}

All of the ULX pulsars so far discovered occupy the same parameter space as HMXBs in the P$_{\mathrm{orb}}$ -- P$_{\mathrm{sup}}$ plane shown in Figure \ref{fig:example_figure}. We suggest that this is strong evidence that the super-orbital periods in ULXs are being driven by the same mechanisms as disc-fed HMXBs. All ULX pulsars in Figure \ref{fig:example_figure} lie close to the best fit to the entire HMXB data-set (black line), so this relationship can be used as a first order approximation of the orbital or super-orbital period in newly discovered ULX pulsars, should the other period be known. It is observed in many other types of interacting binary that disc precession on timescales of $\sim$20 orbital periods is typical, which perhaps explains why the ULX pulsars appear to cover the entire parameter space in Figure \ref{fig:example_figure}. One can combine knowledge of the position of ULX systems in Figure \ref{fig:example_figure} with their positions in the spin-orbit parameter space (or Corbet diagram) of HMXBs. The ULXs in the lower left of Figure \ref{fig:example_figure} sit exactly amongst the RL-filling SGXBs on the Corbet diagram. When considered together, the similarities between short period ULXs and disc-fed SGXBs make it highly probable that these systems are accreting from RL-filling super-giant stars, as suggested by other work in the literature. Even at this early stage, and with little data to compare, the predictive power of this relationship for ULX pulsars is apparent.

Things become a little less clear when considering longer period systems, as these objects are unlikely to be RL-filling. The three luminous BeXRBs in our data-set are all considered to have short orbital periods relative to the rest of the population. This is expected, as it is known that long-period BeXRBs have systematically lower peak X-ray luminosities and so are less likely to be seen as super-Eddington sources. The ULX NGC7793 P13 sits amongst the BeXRBs in Figure \ref{fig:example_figure} and closer to these systems than the SGXBs in the Corbet diagram, yet is known to harbour a super-giant. The orbital period of P13 is exceptionally long for a SGXB, suggesting a different state or evolutionary pathway for this system over classical disc-fed SGXBs. Though the source numbers are low, it may be more difficult to use the P$_{\mathrm{sup}}$ -- P$_{\mathrm{orb}}$ relation to help in the classification of longer period ULX systems. However, these data do show that longer period ULXs may be BeXRBs accreting at super-Eddington rates. This leads naturally to a distinction between short and long period ULXs: short period systems accrete persistently through the RL of a super-giant companion and are consistently emitting above the Eddington limit, whereas long period systems are not RL filling and may occasionally develop a transient accretion disc that allows accretion to exceed the Eddington limit for  short intervals. It seems the donor stars in these systems are a mix of super-giant and main-sequence stars (see Table \ref{tab:example_table} for possible counterparts from the literature).

\subsection{The link to DPVs}

The pale blue line in Figure \ref{fig:example_figure} represents the data gathered by \cite{2003A&A...399L..47M} on DPV systems. These systems form an almost completely continuous data-set with the BeXRBs, with a small break in the trend at around P$_{\mathrm{orb}}$ $\sim$ 20d. Given the similarity in the period correlation in the DPV and BeXRB populations, and the fact that the DPV systems seem to mostly consist of a Be star and a highly evolved primary, we suggest that DPVs may be direct progenitors of BeXRBs.

Many papers discuss the spectroscopic observations of DPVs (e.g. \citealt{2012MNRAS.427..607M}; \citealt{2014A&A...567A.140B}; \citealt{2013MNRAS.428.1594G}), all consisting of $\sim$2\,M$_\odot$ late B/early A-type mass donors (of radius $\sim$9\,R$_\odot$) rapidly transferring material onto $\sim$7--9\,M$_\odot$ early B-type primaries with circumstellar discs. This is remarkably close to what one would expect for progenitors of normal BeXRBs or Be + white dwarf systems. The donors are highly evolved and will, if massive enough to ignite He core burning, progress though a He star phase before becoming a compact object. However, it is unclear how many would become white dwarfs rather than neutron stars. As yet, no Be+WD has been dynamically confirmed, but \cite{2019IAUS..346..143G} discuss the search for BeXRB progenitors using UV spectroscopy of a sample of 264 Be stars. They find 16 of these systems to be Be + He star binary systems. Mass estimates are known in just 6 of the population of Be + He star binaries, and just one of these has a companion mass greater than the Chandrasekhar mass (60 Cygni) and is thus a candidate to go supernova and produce a BeXRB. The systems with masses below the Chandrasekhar limit will evolve into Be+WD binaries, which are expected to exceed the Be+NS binaries in number based on population synthesis modelling (e.g. \citealt{2001A&A...367..848R}).

This hypothesis is also supported by the known distribution of spectral types and stellar masses in BeXRBs, as all are known to host Be stars of spectral type earlier than B3 (e.g. \citealt{2008MNRAS.388.1198M}).  This also matches the known spectral types of Be stars in Be + He star binaries (\citealt{2019IAUS..346..143G} and references therein) and DPVs. The Be stars in DPVs are accreting at high enough rates that they will become more massive by the time the secondary becomes a remnant, perhaps matching the mass range observed in BeXRBs. The few of these systems with secondaries massive enough to ignite He core burning will become Be + He star binaries and may eventually become BeXRBs.

Finally, the mechanism producing the longer period in the DPV systems is also found in BeXRBs. \cite{2015MNRAS.448.1137M} show a schematic of a DPV depicting the locations of a hot spot and bright spot on the Be star disc. The precession of the bright spot on the disc is equivalent to the precession of a density wave that produces the super-orbital periods in BeXRBs.

\section{Conclusions}

We have examined the observational properties of ULX pulsars and NS-HMXBs in the context of their spin, orbital and super-orbital periods and show all ULX pulsars follow a tight linear relationship between their orbital and super-orbital periods. This relationship can be used as a predictive tool in newly discovered ULXs, much as it can in HMXBs. It also provides good evidence that the super-orbital periods produced in ULXs are a result of the same mechanisms at work in HMXBs, and that many ULX pulsars may be HMXBs at the extreme end of the accretion scale.

The position of BeXRBs and the so-called DPV systems on the P$_{\mathrm{orb}}$ -- P$_{\mathrm{sup}}$ plane is a compelling argument that the two populations are linked. The component stars of DPV systems, namely an evolved late-type donor and an early-type Be primary, are remarkably similar to what one would expect of the progenitors of Be+WD systems and BeXRBs. It is possible that the most massive DPVs will evolve into Be + He star binaries, with the most massive of these eventually going supernova and becoming BeXRBs. A normal distribution of He core masses in these systems is likely to reflect the predicted ratio of Be+WD and Be+NS binaries from population synthesis models. The supernova will cause the orbital period of the binary to lengthen and the super-orbital period in the disc to be modulated at a longer period. We suggest this makes DPV systems strong candidates as progenitors of both BeXRBs and Be+WD systems.

\section*{Acknowledgements}

This work began when PAC was visiting UCT with the support of a Leverhulme Emeritus Fellowship.  PAC would also like to thank Linda Schmidtobreick for pointing out the DPV relation during the Vina del Mar symposium in 2011.  LJT acknowledges support from the National Research Foundation of South Africa.




\bibliographystyle{mnras}
\bibliography{ulx-periods} 








\bsp	
\label{lastpage}
\end{document}